\documentclass[11pt]{article}
\usepackage{epsf}
\usepackage{graphicx}        
\input epsf
\def\ll{\label}
\def\re{\ref}
\def\c{\cite}

\def\r1{(\ref{$1})}

\def\ep{\epsilon}

\def\ba{\begin{array}{c}}

\def\ea{\end{array}}

\def\l{\left}
\def\l({\left(}
\def\r){\right)}
\def\r{\right}

 \def\be{\begin{equation}}
\def\bc{\begin{center}}
\def\ec{\end{center}}
\def\bit{\begin{itemize}}
\def\eit{\end{itemize}}
\def\ee{\end{equation}}
\def\ed{\end{document}}
\def\bea{\begin{eqnarray}}
\def\eea{\end{eqnarray}}
\def\efr{\end{flushright}}

\begin{document}
\title{ 
Exact  asymmetric
Skyrmion 
in anisotropic  ferromagnet
 and 
  its helimagnetic application 
 }
\author{ Anjan Kundu\\
 Theory Division,
Saha Institute of Nuclear Physics\\
 Calcutta, INDIA\\
{anjan.kundu@saha.ac.in}}
 \maketitle

{
 05.45.Yv,
02.40.-k,
75.10.Hk
}
\begin{abstract}
 Topological Skyrmions as intricate spin textures  were observed
 experimentally    in  helimagnets on  2d plane. Theoretical foundation of such 
 solitonic states to  appear  in pure  ferromagnetic model,
  as exact solutions expressed through any analytic
 function, was made long ago by Belavin and Polyakov
 (BP). We propose
 an innovative generalization of the BP solution for
  an anisotropic ferromagnet,  based on a physically motivated geometric
(in-)equality,  which takes the exact Skyrmion to a new class
of functions beyond analyticity.
The possibility of  stabilizing  such 
 metastable states  in  helimagnets is discussed with the construction of 
individual Skyrmion, Skyrmion crystal and lattice with asymmetry, likely to be
detected in precision experiments. 

\end{abstract}



\section{ Introduction}

 Unusual spin textures, identified as   magnetic
Skyrmions  of topological origin,
were observed in   helimagnetic crystals like  Mn Si  and  Fe$_{x}$Co$_{1-x}$Si,
in a series of recent experiments
 \c{QHexperim,nutronScattExp,LorentzExp}.  However, the possibility of such
a phenomenon to occur on a two-dimensional (2d) plane in ferromagnetic model
was predicted by Belavin and Polyakov  40 years back, in a pioneering work,
where they found exact Skyrmions in a general framework with topological charge $
Q=N,$ linked to any analytic functions \cite{BP}.  Problem with this beautiful
theoretical result was, that due to scale invariance of the solution it could
give only metastable states, where for stabilizing the Skyrmions one needs
to bring in coupling parameters through  additional interactions.
  In a helimagnetic model a
 Dzyaloshinskii-Moriya (DM) spin-orbital  interaction term is added to the 
spin-spin ferromagnetic interaction, where the DM coupling   breaks the 
  unwanted scale invariance and stabilizes the
solitons  through competing forces between the ferromagnetic  and the effective DM
interactions. Recent experimental observation of Skyrmions in magnetic models
is therefore  a landmark  confirmation of  an old   basic theoretical result
proposed in  \c{BP}.    

It is a bit surprising, that in the  recent large  collection of high-profile
theoretical and experimental
 work dedicated to  the magnetic Skyrmions in helimagnets
\cite{topological,top2,QHexperim,nutronScattExp,LorentzExp}, rarely the
work of BP \cite{BP} is mentioned, which on the other hand is the
theoretical basis for the topological Skyrmion in magnetic model in 2d.  It
is also rather unexpected, that no other extension or generalization of the BP
solutions was proposed  in 2d, over these long years of development in the
subject.  Our motivation here, is to bring
in a novel contribution to the subject, by discovering a family of exact
Skyrmions for an anisotropic extension of the ferromagnetic model, through a
physically motivated geometric (in-)equality, which generalizes the BP
result by going beyond the analytic functions and linking the Skyrmions to
the contemporary $\bar \partial $ problem \cite{dbar}.
Such Skyrmions exhibit more inherent asymmetry and anisotropy, properties  
akin also to the helimagnetic crystals with noninversion symmetry and
anisotropy exhibited in 2d. Therefore, a natural application 
of such asymmetric solitons to the helimagnetic model, through their stabilization 
under DM interaction might  be a promising perspective to be verified in  a precision
experiment.  
\section{ Ferromagnetic and helimagnetic  models and BP solution}

As evident from   the experimental images, the  Skyrmion magnetic pattern in
helimagnets show
  extended structure for  individual Skyrmions  in a  30 nm range,
 slowly varying  over  several lattice
spacings \cite{LorentzExp}. Therefore, 
one can go for the  continuum limit reducing the  ferromagnetic Hamiltonian  $H_f $ 
 from the Heisenberg spin
  model  on a 2d lattice with nearest neighbor interactions as 
\bea H_{HS}=\frac 1 2 \sum_{<{\bf j}, {\bf j'}>} {\bf s_j}{\bf s_{j'}}
\ \longrightarrow  \ \
  H_f= \frac 1 2\int d^2x (\nabla {\bf M})^2, \ 
  \ll{HSC}\eea
{where} $ \ \ {\bf s_j} \to {\bf M}, \ \ {\bf s_j}^2 \to {\bf M}^2=1. $
For constructing a  helimagnetic model a crucial   addition of a 
spin-orbital interaction is needed to the
ferromagnetic Hamiltonian (\ref{HSC}) in the form of  
a DM term    \be 
 H_{dm}=\int d^2x({\bf M } \cdot [\nabla \times {\bf M}])
,\ll{DMc}\ee
which was proposed phenomenologically way back  in another landmark work
\cite{landau}. 

Note, that the DM Hamiltonian (\ref{DMc}) exhibits an
 explicitly broken  space inversion symmetry as well as  anisotropy, due to 
 the appearance of only first order   derivatives  and 
interaction  of the  2d coordinate  space with the  internal spin space,
 rewriting (\ref{DMc}) as 
\bea  H_{dm}&=& \int d^2x  \left(M^1 \partial _y M^3 +M^2 \partial _x M^3
+ M^3( \partial _x M^2- \partial _y M^1)\right)
\ll{HDM2} \eea
Therefore the  basic helimagnetic model,  for investigating the magnetic
Skyrmions
 of contemporary interest,  may be given simply by the model
Hamiltonian  
$ H_{heli}= J H_f+ D \  H_{dm}
$
where $J$ is the spin exchange parameter, $D$ is the spin-orbital 
 coupling  and the ferromagnetic $H_f $ and the DM Hamiltonian $H_{dm} $ may be given by
(\ref{HSC}) and (\ref{HDM2}), respectively.    

\subsection  { Topological charge }

 Here comes the most interesting topological concept, by  noticing  that 
 the magnetic moment field ${\bf M}(\rho, \alpha)$ defined in  a 2d space ($\rho, \alpha $
being the polar coordinates) takes values on a
 2-{\it sphere} ${\bf M} \in S^2_M $. At the same time,  
for finite energy soliton solutions,  ${\bf M} $
must go to a fixed vector at space infinities    ${\bf M}_\infty(\rho \to \infty)
\to (0.0.1). $   This in turn compactifies the 2D space $(\rho, \alpha) \in { R}^2 $ also to
a 2-{\it sphere} $ S^2_r $ (since space-infinities can be added as a  fixed  
north pole to form a sphere), resulting  to a $S^2_r\to S^2_M $ mapping (see
Fig. 1a)
   
\includegraphics
[width=6cm
]
{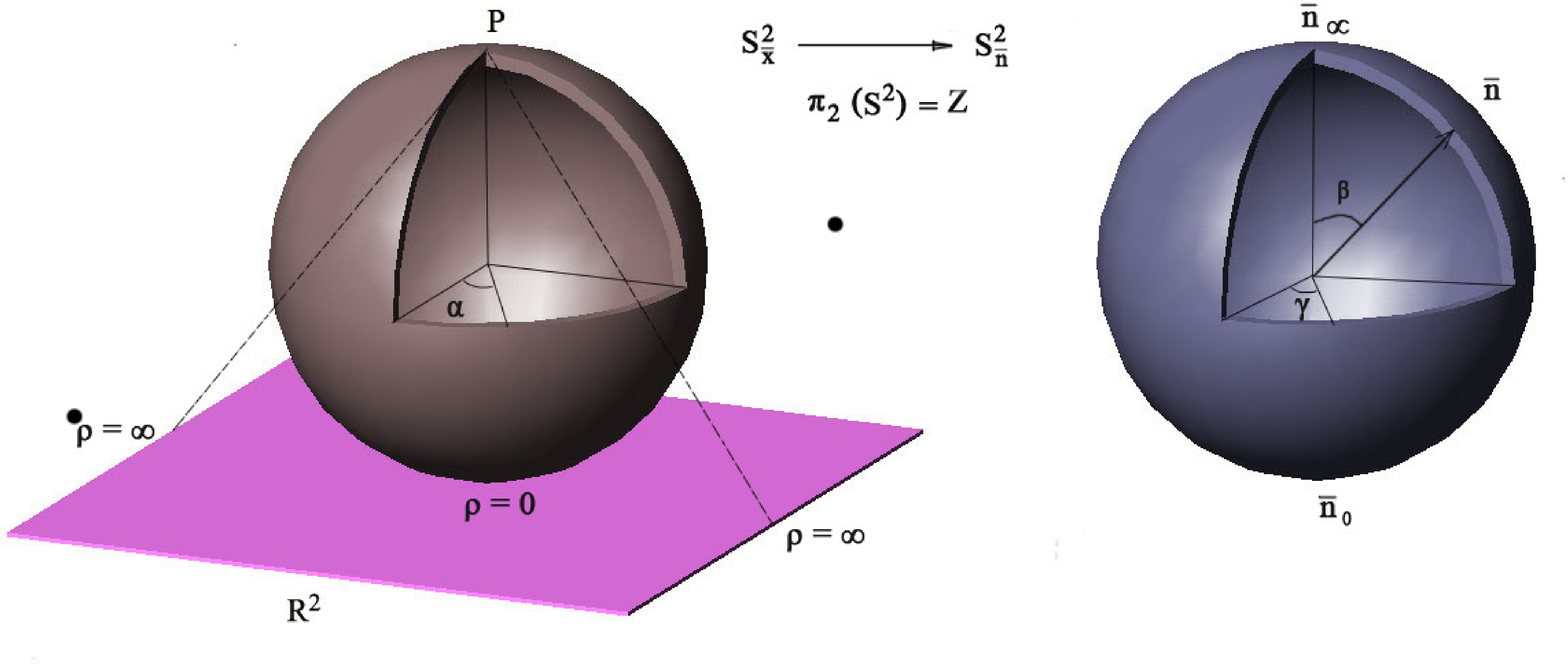} \ \ \ \ \ \qquad \qquad 
\includegraphics
[width=6cm
]
 {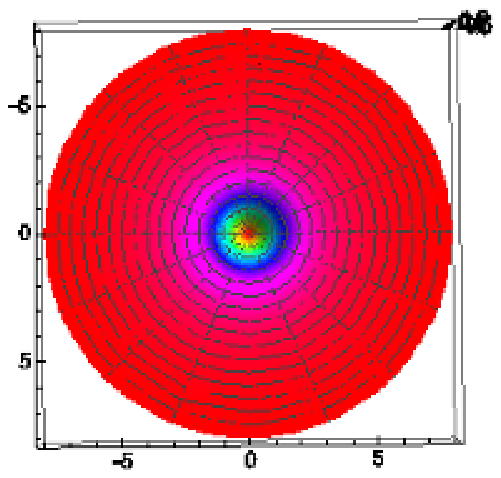}

 \ \ \ \ \ \qquad \qquad a)  \ \ \ \ \ \qquad \qquad  \ \ \ \ \ \ \qquad \qquad 
 \qquad \qquad  \ \ \ \ \ \qquad \qquad
 b)

 \noindent  { Fig. 1:} {\small Topological picture  in 2D field model. \  a)
  {\it sphere to sphere}
mapping with charge $Q$ as the degree of this mapping. b)  
 Symmetric Skyrmion in ferromagnetic model with
topological charge $Q=-1$} 
 
\vskip .7cm

The {\it  degree}
of this mapping, which   counts how many   times 
 Skyrmion  field ${\bf M} $  wraps  the target sphere $S^2_M $,
 when  coordinate space $S^2_r $ is
  covered once, may be defined as the topological charge $Q$. 
 It is evident therefore, that   topological charge is a geometrical property
 that  can take only  integer values  $Q=N$ and  
since it is conserved topologically, Skyrmion with a fixed charge can
not decay into a configuration with a different charge and hence acquires
a topological stability.
\subsection{BP Skyrmions}
The original idea of a topological Skyrmion to appear in pure ferromagnetic
model (\ref{HSC}) was put forward in \cite{BP}.
  However due to scale invariance of  such Skyrmions the states remain
metastable and difficult to observe in experiments. Therefore for stabilizing
the solitons some scale must be introduced,  as done by    
 additional $ D \  H_{dm}$ coupling in a helimagnetic model. 

For understanding  therefore the original idea of \cite{BP} for constructing
the Skyrmions,  
 let us  concentrate on ferromagnetic Hamiltonian $H_f$ (\ref{HSC}) alone
and note that,  since charge $Q$ for a Skyrmion is conserved, if we 
ensure a lower bound for the Hamiltonian  like $H_f \geq \ {\rm const.} |Q| $,
the  {\it energy}  can not decay into the  vacuum,  making the
state
 stable and at the same time 
the  configuration reaching  the lowest bound (i.e. when {\it equality} is
achieved)
 should minimize the energy, producing 
  thus the needed soliton solution. To elaborate the    picture let us 
express magnetic moment ${\bf M}$  through two spherical angles 
$(\beta,\gamma)$  on $S^2_M $ : $$\beta (\rho, \alpha) \in [0,\pi],\gamma 
	 (\rho, \alpha) \in [0,2\pi]$$ as
\be  M^{\pm} = M^1\pm iM^2= \sin \beta e^{ \pm i \gamma}, \  
 M^3=  \cos \beta 
\ll{M123} \ee
   and rewrite   for  convenience  ferromagnetic Hamiltonian (\ref{HSC}) as 
\be  H_{f}= \int d^2x h_f(\beta,\alpha), \ h_f(\beta,\alpha)
= \frac 1 2 ( (\nabla {\beta})^2+\sin ^2 \beta 
(\nabla {\gamma})^2)
\ll{Hf} \ee  and the topological charge as 
\be  Q= \frac 1 {4 \pi} \int  d^2x q(\beta,\alpha), \ \ 
 q(\beta,\alpha) =\sin  \beta [\nabla \beta\times \nabla \gamma ]
 \ll{Q} \ee
For further insight  we { introduce} two   field vectors
\be   {\bf X}= \nabla \beta, \ \ {\bf Y}=\sin \beta \nabla \gamma \ll{XY} \ee
to rewrite 
\bea &&  h_f(\beta,\alpha) = \frac 1 2 ({\bf X}^2+ {\bf Y}^2), \ \
 q(\beta,\alpha)=[{\bf X}\times {\bf Y} ]
\ll{hf} \eea
which by using an obvious  inequality
\be  \frac 1 2 \left((X_1- Y_2)^2+(X_2+Y_1)^2\right) \geq 0  
\ll{CRin}\ee
 and regrouping the terms as  
$$  \frac 1 2 ({\bf X}^2+ {\bf Y}^2) \geq 
(X_1 Y_2-X_2Y_1) $$
 leads clearly to  $h_f\geq |q| $, which by  integrating both sides
 gives the required inequality $H_f \geq 4 \pi  |Q| $.
For answering the  next question, that is,  when  we can get  the {equality}  
 $H_f = 4 \pi  |Q| $, we focus again on  (\ref{CRin}) and see clearly that
equality is achieved 
at the relations  \be X_1= Y_2, \ \ X_2=-Y_1 , \ll{X1Y2}\ee
  which are simple  first order equations, 
allowing exact  Skyrmion
solutions. Interestingly,   mapping ${\bf M}$ to a complex field   (stereographic
projection): 
 $ \ \frac {M^+} {1+M^3}=f(z)=u+iv  \ $ one can 
  map the minimizing condition  (\ref{X1Y2}) to a similar relations 
 $ \partial_1 u= \partial_2 v, \ \ \partial_2 u=- \partial_1 v, $
 which is known to us very well as the  Cauchy-Riemann (CR) condition for
 analyticity  $\partial_{\bar z}f(z)=0 $ of a complex 
function $f(z). $   
Therefore the BP result concludes, that 
 any analytic function
\be  f(z)= \frac {\prod_i(z-z_i)^{n_i} } {\prod_j(z-z_j)^{n_j}}, \
N=\sum_i
n_i-\sum_jn_j
 , \ll{BPsol} \ee
  with $z_i,z_j $ any arbitrary constant complex numbers,
 would be an exact Skyrmion solution of the ferromagnetic model with
topological charge $
Q= N$, where we have explicit  scale invariance $z \to z_0 z  $ of the
solution.

{ The  simplest exact }  solution   obtained from the above
general result as 
  $f(z)= \frac 1 z = \frac 1 \rho e^{-i \alpha},  $ leading to the  Skyrmion 
solution for the magnetic moment in the ferromagnetic model (\ref{hf})
  as \be  M^3=\cos \beta= \frac {\rho^2 -1} {\rho^2 +1},
 \ \ M^3 (\rho=0)=-1, 
\ \ M^3 (\rho=\infty)=+1,  \ \gamma= -\alpha, \ll{BP1} \ee
 shows perfect 
  circular symmetry (see Fig. 1b).

 It is rather surprising, that even for the helimagnetic
model exhibiting  high level of asymmetry and anisotropy, the contemporary theoretical models 
look for only Skyrmions with circular symmetry. Our findings, as we show
below, will be tuned towards asymmetric Skyrmion solutions.

\section {Beyond BP Skyrmions}
  Our motivation is to 
 look for more general solutions beyond symmetric Skyrmions, which might be detected
in  {future}  precision experiments.
  We intend to find some way to extend the BP solutions. However
since the BP procedure  
seems to be built on a  specific   construction  using  a particular
inequality   (\ref{CRin}), it is
difficult to get any clue for its generalization.
Therefore, we turn for  new ideas to some physically motivated inequalities 
of geometric nature with an universal appeal, called  isoperimetric
inequalities.

\subsection{ Isoperimetric
inequalities}
It is a common  experience, that the soap bubbles, deformed at its starting
always try to take the spherical shape in time. This is because the bubble surface $S$ is
lower bounded by its volume $V$ and tends to reach the equilibrium point 
by minimizing its surface area, which is achieved for a sphere, at reaching the
equality. 

Interestingly, this universal concept remains   valid also in 2d,
where for any  closed curve its perimeter $P$ is bounded from below by the 
area enclosed: $A$, through the relation  $P^2\geq 4 \pi A $.
It is not difficult to check, that the equality is reached 
 at the maximum symmetry, i.e. when the 
  the curve turns into a circle.
  
This picture becomes even more interesting at discrete symmetries, where for
a polygon of $n$-sides the perimeter $P_n $ and the enclosed area $A_n$ are
related through an inequality  
 $ P_n^2\geq c_n A_n, \ \ c_n=(4n\tan \frac \pi n) .$ The equality is
achieved, as expected, at the
 symmetric situation,  when the polygon becomes a regular one with equal
sides and equal angles (which is a nice check  using simple geometric
relations).   

Particular cases of this (in-)equality with $ n=4,3$ will proved to be
important for us.   

\subsection{Inequalities for parallelogram and triangle}
First let us consider a parallelogram 
generated by two vectors   $ {\bf X} $ and  $ {\bf Y},$ which should
satisfy the  inequality  for n-polygon with $n=4 $:  $P_4^2\geq 16 A_4  $, since  
 $c_4= (4 \cdot 4\tan \frac \pi 4)=16 $. The inequality expressed through these
vectors therefore takes the form (see Fig. 2a) $$  
(2(|{\bf X} | + | {\bf Y}|))^2 \geq 16 |[{\bf X}
\times {\bf Y} ]|, $$
where  perimeter $P_4 $ and  area $A_4$  of the 
 parallelogram
 are  rewritten through the  generating vectors.
Using further a  simple relation, known as Schwartz inequality: ${\bf X} ^2 +  {\bf Y}^2
\geq 2 ({\bf X} \cdot {\bf Y}) $ we derive  an important  relation   
\be  \frac 1 2 ({\bf X} ^2 +  {\bf Y}^2)
 \geq |[{\bf X}
\times {\bf Y} ]|,  \ll{paralel} \ee which surprisingly coincides 
 with the ferromagnetic Hamiltonian
and charge densities (\ref{hf}) if we represent our vectors as (\re{XY}),
yielding the same relation $h_f\geq q $  or its integrated version $ H_f\geq 4 \pi Q
$. Thus we could derive the same lower bound of the ferromagnetic
Hamiltonian $H_f$ through  topological charge $Q$, as obtained by BP \cite{BP}
 using a restricted inequality  (\re{CRin}), from a general  framework
 of universal inequalities  for polygons through a
geometric construction. Now, let us see how further we can go in retrieving 
 the BP construction from the present  geometric approach. As we
should have guessed, the equality in (\ref{paralel}) for the parallelogram    
 would be achieved for its maximum symmetry, i.e. when it reduces to a
square having (see Fig. 2a)  \be 
|{\bf X} | = | {\bf Y}|, \ \ \mbox{and} \ \ 
({\bf X} \cdot {\bf Y})=0, \ \ \mbox{with } \ \angle ({\bf X}, {\bf Y})
\equiv \theta_0=90^o . 
 \ll{square} \ee
It is easy to check this intriguing fact, that a   geometric equality,
 obtained for a square (\ref{square})  from an overall symmetry  is
exactly compatible with the self-duality condition (\re{X1Y2}) linked to the
CR condition $\partial_{\bar z}f(z)=0, $ for an analytic function giving
 exact Skyrmion solutions in  the ferromagnetic model \cite{BP}.  
The  relation like   (\ref{square})   found here linked to the  CR condition  
can give also an  uncommon realization for the analyticity condition of a
complex field $f=u+iv $ in the form 
\be |\nabla u | = |\nabla v |, \ \ \ 
(\nabla u \cdot \nabla u)=|\nabla u |^2 \ \cos 90^o=
0, \ \  \ \angle (\nabla u, \nabla v)
=90^o
 \ll{CRnew} \ee
\section{Anisotropic ferromagnet and extension of BP result}
Inspired by the above success in rederiving the exact BP Skyrmions for the
ferromagnetic model, following a different geometric route of universal
nature related to a parallelogram, we intend to focus now 
 on  a triangular inequality,  in the hope for
generalizing the BP result.
Therefore, 
we  focus on the   above polygon relation and specialize for a new  case:    $n=3$
leading to the inequality $ P_3^2\geq c_3 A_3, \ c_3=(4\cdot  3\tan \frac \pi 3)
= 12 \sqrt{3}$ .  The triangle generated by the above two vectors 
clearly  defines the   perimeter of the triangle 
 as $P_3=|{\bf X} | + | {\bf Y}|+ |{\bf X}- {\bf Y }| $
while the area of the triangle may be given through 
$A_3= \frac 1 2  |[{\bf X}
\times {\bf Y} ]| $
expressing the triangular inequality   explicitly as (see Fig. 2b)
\be 
(|{\bf X} | + | {\bf Y}|+ |{\bf X}- {\bf Y }|)^2 \geq 6 \sqrt{3} |[{\bf X}
\times {\bf Y} ]|, 
 \ll{triang1} \ee 

\includegraphics
[width=6cm
]
{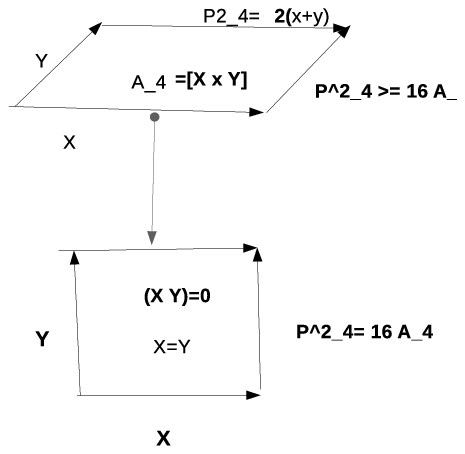} \ \ \ \ \quad  \qquad \includegraphics
[width=6cm
]
{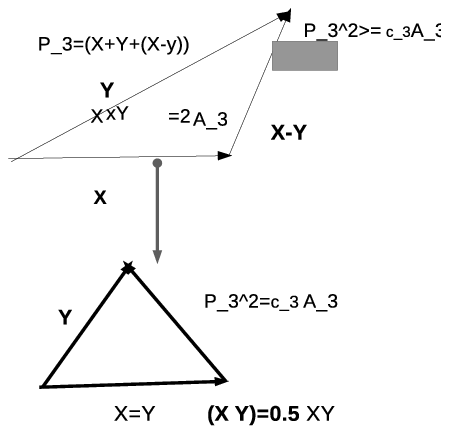} 

  \ \ \ \ \quad  \qquad a)  \ \ \ \ \quad  \qquad \ \ \ \ \quad  \qquad 
\ \ \ \ \quad  \qquad \ \ \ \ \quad  \qquad b)

\noindent   { Fig. 2:} {\small  Isoperimetric (in-)equalities. a) 
Parallelogram inequality reaching equality at {\it square} linked to $H_f\geq Q $ for ferromagnet
 and BP Skyrmion related to analytic functions. b) Triangle inequality
reaching  equality at {\it
equilateral} triangle, linked to new anisotropic  ferromagnet ${\tilde H_f}\geq Q $ 
 and asymmetric Skyrmion related to $\bar \partial $-problem functions. }
\vskip .7cm

Further working on the LHS and using  repeatedly the Schwartz inequality one
derives from (\ref{triang1}) a new relation   
\be 
\frac 1 2 ({\bf X}^2  +  {\bf Y}^2-({\bf X} \cdot {\bf Y}))  \geq \frac
{\sqrt{3}}
2  |[{\bf X}
\times {\bf Y} ]|, 
 \ll{triang2} \ee 
which gives a physical interpretation for an extended  ferromagnetic
model. To see this we represent the vectors as (\ref{XY}) and comparing with
the expression (\ref{hf} used for the ferromagnetic model in the BP
case, define a new  anisotropic ferromagnetic model 
as 
\bea 
& & {\tilde  H}_f= H_f+
H_{anis}, \ \  \nonumber  \\ & &  H_{anis}= \frac 1 2 \int d^2x \frac 1 { (1-{M^3}^2)}
\left ({M^2}{\nabla M^1}- {M^1}{\nabla M^2})\cdot \nabla M^3\right ) = - 
\frac 1 2 \int d^2x\sin \beta (\nabla \beta \cdot \nabla \gamma ), 
\ll{Hfanis} \eea 
where we have expressed the vectors   ${\bf X}, {\bf Y } $ through
the  Magnetic moment vector ${\bf M} $ in deriving the anisotropic part $H_{anis} $.
Interestingly, the   lower bound for this  new  anisotropic ferromagnetic Hamiltonian
${\tilde H}_f $  through topological
charge $Q$  follows from 
 the relation (\ref{triang2}) as
\be
 {\tilde H}_f \geq 2 \pi \sqrt{3}  Q ,
\ll{AFQ}\ee
where ${\tilde H}_f $ is given by (\ref{Hfanis}) with the standard
ferromagnetic Hamiltonian as in (\ref{HSC}) and the topological charge $Q$
as in (\ref{Q}). We may note by passing, that 
comparing with the lower bound for the ferromagnetic model:  $H_f \geq 4 \pi  |Q|  $  
 as appears  in the BP construction,
 the bound  (\ref{AFQ})  we
get  here  for the anisotropic extension is a bit   lower.

The next important question is, can one find  exact Skyrmion solutions
minimizing the Hamiltonian of the new model (\ref{Hfanis}). Since such a
configuration should saturate  the lower bound in  (\ref{AFQ}) as ${\tilde H}_f =  2 \pi \sqrt{3}
Q$,
we seek for the equality in the triangular relations (\ref{triang1}) or
(\ref{triang2})
which, as per the general philosophy we have witnessed above, should be achieved at a maximal
symmetry, i.e. for an  equilateral triangle  giving the  condition (see Fig.
2b) \be   
   {\bf X}^2= 
{\bf Y}^2 , \ \   ({\bf X}\cdot {\bf Y})=\frac 1 2 |{\bf X}||{\bf Y}| , \  \theta_0=
60^o, 
\ll{60}\ee  which   is consistent   with  the relation   \be   
 X_1=\cos \theta_0 Y_1+\sin \theta_0 Y_2, \  X_2=- \sin \theta_0 Y_1+ \cos \theta_0
Y_2, \  \theta_0= 60^o \ll{XYanis}\ee
and clearly is  a generalization of (\ref{X1Y2}).
The   relation
(\ref{XYanis}) or equivalently the  equilateral condition (\ref{60}) 
 minimizes the model Hamiltonian ${\tilde H}_f $ and therefore 
take us beyond the BP Skyrmion.  At the same time, the minimization condition 
(\ref{XYanis}) maps  under the stereographic projection to a  
 complex function  generalizing an analytic function 
and hence satisfying an intriguing extension of  the  
CR-condition:  
\bea  & & F(z, \bar z)=u+i v , \  \ \partial_{\bar z}F(z, \bar z) \neq 0  \nonumber \\
& & \partial_1 u=\cos \theta_0  \partial_1 v +\sin \theta_0  \partial_2 v, \
 \partial_2  u=- \sin \theta_0  \partial_1 v+ \cos \theta_0  \partial_2 v, \  
 \ll{CRK} \eea 
with $\theta_0=\frac \pi 3 $ in the present case. This is  like a  nontrivial rotation
 in the 2d gradient space $\nabla
u=\hat R_{ \theta_0 } \nabla v
, $  where
rotational angle $ \theta_0= \frac \pi 2$  recovers the standard CR
relations,
 while  $ \theta_0= \frac \pi 3$ corresponds to our case, giving  new exact Skyrmion solutions
through a  class of complex functions beyond usual   analyticity.

In general for finding the topological solitons in such magnetic models 
for the magnetic moments
 $$ M^3=\cos \beta (\rho, \alpha), M^\pm = \sin \beta e^{\pm i \gamma(\rho, \alpha)}
 $$ one has to solve the
governing  Euler-equations for the angles  $\beta, \gamma $ by
minimizing the model Hamiltonian.
 However, these equations are highly { nonlinear coupled 
PDEs} and    difficult to solve in general.`
  It is therefore a pleasant rescue  in the present case, that 
 the configuration satisfying the energy minimizing  condition (\ref{XYanis}), which can be expressed
 explicitly as
 \bea 
 \partial_\rho \beta =\frac 1 2 \sin \beta ( \partial_\rho  \gamma  + \frac
 {\sqrt 3}
  \rho \partial_\alpha  {\gamma} ) ,\  \ \
\frac 1 \rho \partial_\alpha  {\beta}   =
\frac 1 2 \sin \beta ( -\sqrt 3 \partial_\rho  \gamma  +
\frac 1 \rho \partial_\alpha \gamma) ,
 \ll{SDK} \eea 
corresponds automatically  to the solutions of the
Euler equations derived from our Hamiltonian ${\tilde H}_f $ (\ref{Hfanis}). Thus we can
solve the   governing Euler  equations of the model, without really  solving them
directly, but by solving only much simpler first-order equations
(\re{SDK}), which 
fortunately, allows exact Skyrmion solutions in a general form.
\subsection{ General Skyrmion solution  in anisotropic ferromagnet through
generalized analytic function}
Recall that the general BP Skyrmions for the ferromagnetic model
(\ref{Hf}) can be given by any analytic function (\ref{BPsol}), which may
be expressed by 
\be f(z)=|f
 | (\cos \theta_f +i \sin \theta_f) , \mbox{ with}\  \ \partial _{\bar z}f(z)=0
.\ll{f} \ee
 On the other hand,   
for our anisotropic ferromagnetic model
(\ref{Hfanis}) the solutions for  the energy minimizing
condition (\ref{SDK}) consistent with the {\it generalized} CR relations
(\ref{CRK}), 
 may be given by  exact   Skyrmions  linked to any complex function $F$ of the
form
\bea & &  F(z, \bar z)=  - \bar c_1 f(z) +c_2 \bar f( z), \nonumber \\
& &  c_1= i-e^{-i \theta_0}, \ c_2= i+e^{-i \theta_0}, \  \mbox{with }\
\theta_0=60^o, \ll{F} \eea
showing  $\partial _{\bar z} F(z, \bar z)=c_2 \partial _{\bar z} \bar f( z) \neq
0, $
 where $f(z) $ is any analytic function in the  form (\ref{BPsol}) and
$\bar f(z) $ is its complex conjugate. Note, that since in  (\ref{F})
 one
gets $c_1=2i, c_2=0 $, for 
$ \theta_0=90^0$ ,   function $F $  reduces to an analytic function $2i f(z) $, as
expected for a BP Skyrmion.

It is rather surprising, that a restricted class of complex functions, which
goes beyond the analyticity with $\partial _{\bar z} F \neq 0$, appears
here as exact Skyrmion solutions with high degree of asymmetry. To analyze
the property of such functions, which satisfy a special  condition
$ (c_1 \partial _{\bar z}F + c_2\partial _{\bar z} \bar F)=0   $ together with its complex
conjugate, 
 and to see  how they differ from the
$\bar \partial $ problem  known  in the literature \cite{dbar}, we represent
the function in the allowed integral form \cite{dbar}
\bea 
 F(z, \bar z)&=& -\bar c_1 f(z)-\frac 1 {2\pi i} \int \int_\Omega \frac {\partial _{\bar
\zeta}F(\zeta, \bar
\zeta)} {(\zeta-z )} d \zeta \wedge  d \bar\zeta 
\nonumber
 \\ \
&=& -\bar c_1 f(z)-c_2\frac 1 {2\pi i} \int d \bar\zeta   {\partial _{\bar
\zeta} \bar f(\zeta)}  \int d  \zeta \frac 1 
 {(\zeta-z )}     \ \
= -\bar c_1 f(z)+c_2 \bar f(z)
\ll{Fint} \eea
which recovers the solution through $\bar \partial $ like problem, though it differs a
bit from the usual $\bar \partial $ assumption of $\partial _{\bar z}F =f(z)$ \c{dbar}.

We can therefore find the  
  exact   Skyrmion solutions for the  present model 
 in a general form for the fields $\beta, \gamma $ as
 
\bea  \tan \frac \beta 2 e^{i \gamma}=F\equiv  |f |(\sin( \theta_f+
\theta_0) +i \sin \theta_f), \ \theta_0=60^o \ll{aSkyrmGen} \eea
  given explicitly as
\bea    \beta = 2 \tan^{-1}( |f | \delta_f ), \ \delta_f^2= \sin^2( \theta_f+
60^o) + \sin^2 \theta_f), \noindent \\
\sin \gamma= \frac 1 {\delta_f} \sin \theta_f, \ \
 \cos \gamma= \frac 1 {\delta_f} \sin (\theta_f+60^o)  \ll{asbetgamma} \eea
Note that with the analytic function f(z) (\ref{f}) given as 
(\ref{BPsol}) the  localized Skyrmion solution for our anisotropic
ferromagnetic model  can be given in the general form  (\ref{asbetgamma})
with topological charge $Q=\pm N$.  

For understanding the structure of such topological solitons    
let us focus more closely on the
 simplest case with $Q=-1$,  obtained as a reduction of
(\ref{asbetgamma}) with 
$  |f |= \frac 1 \rho,$ and $ \ \ \theta_f= -\alpha$ 
to construct the explicit exact Skyrmion (see Fig. 3a for graphical
representation)  
\bea & & M^3=\cos \beta= 
\frac {{\rho }^2 -\delta^2_f (\alpha)}{{\rho }^2 +\delta^2_f (\alpha)} , \ \
\delta^2_f (\alpha)=\sin^2\alpha + \sin^2(\alpha -\theta_0)\nonumber \\
& & \cos \gamma = -\frac 1 {\delta_f} {\sin (\alpha -\theta_0)} , \
\theta_0=60^o ,  \ Q=-1 
 \ll{K1skyrm}\eea

\includegraphics
[width=6cm
]
{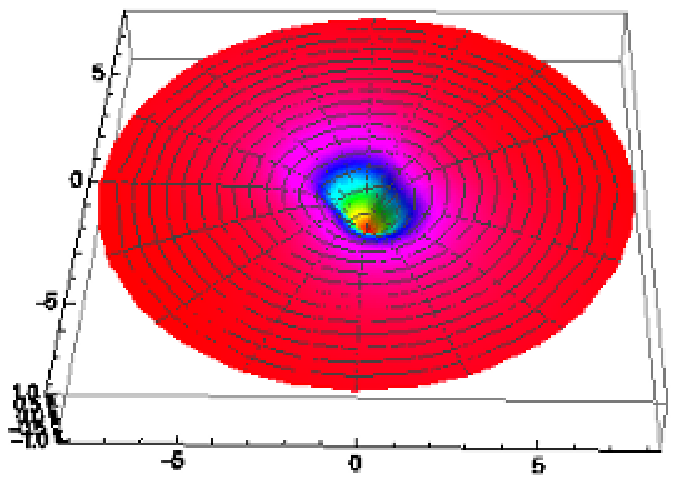} \ \ \ \ \quad  \qquad \includegraphics
[width=6cm
]
{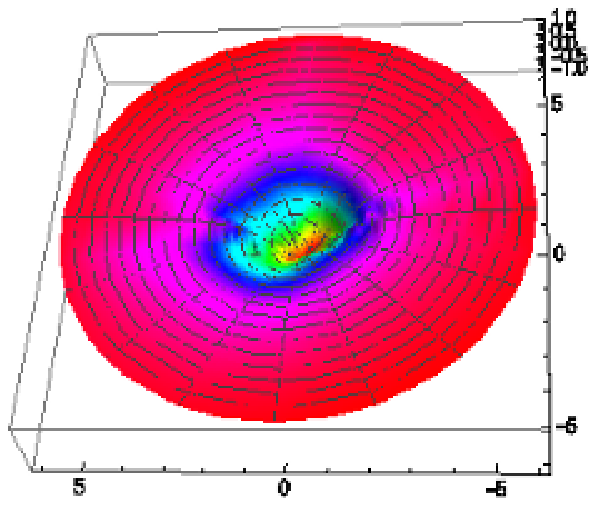} 
  
\ \ \ \ \quad  \qquad a)  \ \ \ \ \quad  \qquad \ \ \ \ \quad  \qquad 
 \ \ \ \ \quad  \qquad \ \ \ \ \quad  \qquad b)

\noindent   { Fig. 3:}{\small Asymmetric Skyrmion for $M^3$ in magnetic models.
  a)  Exact asymmetric Skyrmion   in
anisotropic ferromagnetic model with $Q=-1. $
 b)  Asymmetric Skyrmion  in Helimagnetic model, with  soliton a) deformed 
under  perturbation of    DM coupling.}
\vskip .7cm 

It is important to note that even in the simplest case the exact Skyrmion
solution does not exhibit circular symmetry since $\beta=\beta(\rho, \alpha) $
(Fig. 2a), whereas at  
  $\theta_0=90^o $ , we get  $  \  \delta_f =1$ and hence the 
 asymmetric Skyrmion reduces to the BP case with circular symmetry:
 $ \beta=\beta(\rho), \gamma
=-\alpha.$  Note that  the Skyrmions in  both these cases 
 are localized structures with the
asymptotic:  
$M^3|_{\rho =0}=-1, \ M^3|_{\rho
=\infty}=1 
$

The simplest asymmetric  topological soliton
constructed here, is
perhaps  the only known    spherically nonsymmetric exact solution   found in any  solvable
model,
in two and higher dimensions. We hope therefore, that the present novel approach
could be applicable to other known models for discovering new asymmetric
solutions.

\section{Asymmetric Skyrmions in Helimagnetic model}
Our next intention is to apply our result for seeking asymmetric Skyrmions for the
helimagnetic model, with the hope that they could be detected in future
precision experiments.    
Note that in spite of the discovery of exact Skyrme solitons with $Q=\pm N $ in a general
form (\ref{F}),
for anisotropic ferromagnetic model (\ref{Hfanis}), one should pay attention
that such solutions are again scale invariant as  the BP Skyrmions and
hence difficult to observe in real experiments due to their metasability.
Therefore for physically observed configurations such Skyrmions need to be
stabilized by coupling to  other interactions, which can break the
explicit scale invariance of the system. This was realized in the known  helimagnetic
model \cite{top2}, where a DM interaction (\ref{HDM2}) was considered in addition to the
ferromagnetic Hamiltonian  (\ref{Hf}).
We shall adopt a similar strategy in stabilizing the asymmetric Skyrmion 
solutions of our 
anisotropic ferromagnetic model (\ref{Hfanis}) by adding 
 the same DM interaction (\ref{HDM2}) to it.
 Therefore,
we propose to look for the Skyrmion spin pattern in an anisotropic
helimagnetic model, theoretically, by taking the helimagnetic model as
\be H_{anheli}=
\frac J 2 (H_{f}+H_{anis}) +D H_{dm} \ll{anheli} \ee
where the ferromagnetic Hamiltonian $H_f $ is given  in the standard form
(\ref{HSC}),
$ H_{anis}$ is its anisotropic extension as in (\ref{Hfanis}) and $H_{dm} $ is
the DM interaction term (\ref{HDM2}).
Since helimagnetic materials show a high degree of asymmetry, 
noninversion symmetry and additionally anisotropy, when projected on a 2d
plane, it might be reasonable to consider a  helimagnetic model like
(\ref{anheli}) showing explicit asymmetry and anisotropy.

Recall, that the DM interactions in magnetic crystals in
general  might have
antisymmetric as well as symmetric contributions under spin exchange,
intensity of which would depend on the crystal symmetry \cite{Moriya}. Notice that, the
$H_{dm}$ term
 containing a single  derivative is antisymmetric under space inversion,
representing the antisymmetric part of the DM interaction, while $H_{anis}$
part involving  two gradients is symmetric under space inversion and might
appear in magnetic crystals with higher symmetry.


  For finding the spin texture in this helimagnetic model (\ref{anheli}) described by 
the    Skyrmion solution, one has to
solve 
  the  associated  Euler equations for the two angle variables
$\beta(\rho,\alpha),$ and $ \ \gamma (\rho,\alpha), $ 
 derived from the  model
Hamiltonian.
The corresponding energy
minimizing  
Euler equations in the static case for our helimagnetic model  (\ref{anheli}) may be given by 

\bea
 && J \ \left (\nabla^2\beta - \frac 1 2  \sin 2 \beta (\nabla
 \gamma)^2
 -\frac 1 2 \cos \beta  (\nabla\beta \cdot \nabla \gamma)\right)+ 2 D \  
M_\beta (\beta,\gamma) =0 , 
\ll{betaEq} 
 \eea for angle $\beta(\rho,\alpha) $ and   
\bea
 && J \ \left(\sin 2 \beta (\nabla\beta \cdot \nabla \gamma) +
\sin^ 2 \beta (\nabla^2 \gamma)
 -\frac 1 2( \cos \beta (\nabla\beta)^2+\sin \beta \nabla^2\beta 
\right) +2 D \  M_\gamma
(\beta,\gamma)=0,  \ll{gammaEq} 
\eea
 with respect to angle $\gamma(\rho,\alpha) $ , where   
the additional  DM terms are

\bea
&& M_\beta (\beta,\gamma) \equiv
 \ \sin^ 2 \beta( \partial_\rho \gamma \cos (\gamma-\alpha) + \frac {1}
{\rho }
\partial_\alpha \gamma \sin (\gamma-\alpha))   \ll{EqDMbeta}
\eea and \bea
&& M_\gamma (\beta,\gamma) \equiv 
- \ \sin^ 2 \beta( \partial_\rho \beta \cos (\gamma-\alpha)+  \frac {1}
 {\rho } 
\partial_\alpha\beta \sin (\gamma-\alpha)
. \ll{EqDMgamma}
\eea
It is crucial to observe, that under   a scale transformation: 
  $\rho \to  \lambda\ \rho  $ the
exchange terms and the DM terms behave differently, since all exchange terms
contain gradients  in second order, while the terms in the DM interaction
have them only in the first order. As a result     under the above  scale
transformation   only the DM terms (\ref {EqDMbeta}, \ref{EqDMgamma})
acquire a multiplicative factor $\lambda $ while the exchange terms remain
scale invariant. Therefore, in the absence of  DM interaction the scaling
factor $\lambda $ remains arbitrary including a vanishing value, making the
ferromagnetic Skyrmion to be metastable.    
However, with the additional  DM terms as in our case,, with the factor $\lambda $ explicitly
appearing in the  Euler equations, it gets fixed through the
coupling parameters  $J,D $ of the model acquiring a nonvanishing value.
Therefore the Skyrmion solutions to the above  Euler equations 
minimizing the helimagnetic Hamiltonian (\ref{anheli}) is likely to be a stable
solution, similar to those in the  known helimagnetic model, solved
numerically and observed  experimentally \cite{LorentzExp}.

These equations  however are highly nonlinear coupled partial differential
equations in two variables and do not allow  in general
separation  of variables due to high degree of asymmetry and therefore
difficult to handle in general. It is also evident that  
 the exact solvability of the ferromagnetic part (\ref{Hfanis}) of
the model is lost now, due to the additional DM interaction with  $D \neq 0 $.
Therefore, for constructing  Skyrmion solutions for the anisotropic 
 helimagnetic model our strategy would be to    treat the system
perturbatively  by  considering  $ \ep=D/J, $ the relative coupling between
the DM and the ferromagnetic exchange interaction, as
   small parameter.  The
perturbative solution  therefore  
may be given by expanding around the arbitrary exact solutions of our
ferromagnetic model:
 \be \beta (\rho,\alpha)=\beta_0 (\rho,\alpha)+  \ep \
\beta_1 (\rho,\alpha), \ \gamma (\rho,\alpha)=\gamma_0 (\rho,\alpha)+  \ep \
\gamma_1 (\rho,\alpha), \ll{Pbetgam} \ee 
in the first order of approximation,  where $\beta_0 (\rho,\alpha), \gamma_0 (\rho,\alpha)
 $ are the  unperturbed solutions  induced by  the anisotropic ferromagnetic  Hamiltonian
(\ref{Hfanis}) alone, while $\beta_1 (\rho,\alpha), \gamma_1 (\rho,\alpha)
 $ are the  deformations suffered, when the DM interaction  (\re{HDM2}) is switched on,
perturbatively. The parameter $D $  also serves as the scaling parameter, breaking the scale
invariance of the unperturbed 
Skyrmions and thus providing the required stability to the soliton solutions.
 Since, we have derived already the set of general  solutions
$\beta_0 , \gamma_0$ in the analytic form (\ref{F}), 
we may  extract the deforming
solutions $\beta_1 , \gamma_1$ by solving the corresponding  
 Euler equations  for the helimagnetic model (\ref{anheli})
through the perturbative expansion (\ref{Pbetgam}). This would reduce the
equations to a linear set of coupled partial differential 
equations, which can be
solved  numerically.  Adding these  perturbing 
solutions  to the  Skyrmions $\beta_0 , \gamma_0$
found above for the anisotropic ferromagnetic model, 
 we can finally obtain the Skyrmions for the helimagnetic model in the form (\re
{Pbetgam}). 

For demonstrating our approach we extract an individual Skyrmions for the Helimagnetic
model   (\ref{anheli})
 taking the unperturbed solitons as the simplest solution (\re{K1skyrm})
shown in Fig. 3a. 
The corresponding  perturbative solution for the Helimagnetic model solved
numerically (see the Supplementary material for the Mathematica 10 code) is
shown in Fig. 3b.
It is interesting to notice, that the circular asymmetry of the
 exact Skyrmion of the anisotropic ferromagnetic model
 has been much enhanced in the helimagnetic
model. Such asymmetric solutions never  predicted earlier for the
helimagnetic models, would encourage to
look for them in future experiments.

Note that one can adopt the above approach without including the 
$H_{anis} $ interaction  term in (\ref{anheli}), which 
 has been considered recently by us \cite{kundu15b} for
finding more general Skyrme solutions in helimagnetic model. However our
present aim is to
propose a helimagnetic model with inherently  asymmetric Skyrmion
solutions which   can not be reduced to a symmetric one and 
might carry experimental relevance.
Our basic solutions are also novel exact  Skyrmions without circular
symmetry (\ref{F}). Note that  exact Skyrmion solution with unit topological
charge without 
spherical symmetry has never   been found earlier.

It is also a matter of concern, that though experimentally the images of
Skyrmion lattice and   Skyrmion crystals in the  hexagonal form have been observed
vividly \cite{LorentzExp}, no satisfactory theoretical proposal seem to has
been put forward, where such crystalline structures 
 could be derived as a solution  from the governing
Euler equations of the helimagnetic model, without neglecting the nonlinear interactions
between the individual  Skyrme molecules, forming the crystal.   
Note, that our approach as outlined above, could give one of such solutions
for the Skyrmion crystals  or Skyrmion lattice in a
helimagnetic model, by perturbing a ferromagnetic  Skyrmion crystal or lattice  with
anisotropy.

Note, that  hexagonal  Skyrmion  crystals  in helimagnets are formed  
from $N=7 $ individual  Skyrme molecules and  the  
Skyrmion  crystals in tern serve as  the building blocks of the 
Skyrmion  lattice, which are formed  through  repeated entries of the   
 Skyrmion  crystals by periodic shifting of their centers.
 Therefore for finding first the  Skyrmion   crystal solution for
our helimagnetic model,
we could proceed   with the  building of   a hexagonal Skyrmion  crystal  for our
anisotropic ferromagnetic model, 
 which can be constructed from the general Skyrmion solutions  in
the analytic form (\ref{aSkyrmGen}), where the  function $F(z, \bar z)$ (\ref{F})
expressed through analytic function $f(z)$ (\ref{f}) may be chosen as a  Skyrme
soliton in  a hexagonal lattice form with
equal sides $a_s, $  explicitly as 
 \be f^{-1 }(z)= \prod_{j=1}^7  {(z-z_j)}= - a_s^6 \rho 
 e^{i \alpha }+\rho^7  e^{i 7 \alpha }
,\ll{f7} \ee
 having  topological charge $Q=-7 $
  with $7$
centers chosen equidistantly by properly adjusting the constant parameters $z_j
$. For finding the corresponding Skyrmion crystal solution for the
helimagnetic model the above crystal solution  could  be subjected to DM-interaction
perturbatively, 

For extending the solution for building the Skyrmion lattice, 
   solution (\ref{f7}) can be repeated
$n$-times with arbitrary $n$ and by shifting the parameters $z_j$,
 which would  again be an exact solution, since the $n$-product of
 solution (\ref{f7})  is another solution, due to general form of the
analytic function (\ref{BPsol}).
 Such Skyrmion lattice subsequently can be treated perturbatively for  DM-interaction
 to get finally the required Skyrme lattice solution  for the
anisotropic 
helimagnetic model (\ref{anheli}).


The  derivation and construction of the  Skyrme crystal and
Skyrme lattice solutions 
 for the helimagnetic model outlined above,
 will not be presented here in detail, since our basic aim is to put forward
our novel idea and to show how it works.

\section{Concluding remarks}

The recent experimental observation of   Skyrmions in helimagnetic materials
on a 2d plane, is a triumph of the theoretical prediction of such exact topological
solitons in a ferromagnetic model proposed 40 years ago by Belavin and
Polyakov (BP) \c{BP},
 though due to the metastable states 
they could not be detected in the original model. On the other hand, during all these
years,  there was  hardly any  proposal  which could generalize the
idea of BP for constructing  novel exact solitons of topological origin in
other 2d models, which consequently, could be tested experimentally.
 We have presented here such a
new idea based on some geometric inequalities of universal nature and
found new exact topological solitons with intrinsic asymmetry in an
anisotropic ferromagnetic model.
This result generalizes the well known Cauchy-Riemann condition as a new
minimization condition for the model Hamiltonian and extends the  BP
Skyrmions 
 to a novel  class of
exact asymmetric Skyrmions. Such Skyrmion solutions are linked to a 
 new class of {\it generalized} functions
related to the $\bar \partial $ problem and might have far reaching consequences in
terms of generalized  symmetries in conformal field theory and related algebraic structures
\cite {ginsperg}. On the other hand, our result could be applied to the
helimagnetic models of contemporary interest, for finding individual
Skyrmions as well as  Skyrmion crystals and lattice  exhibiting asymmetry and anisotropy, as a natural
solutions of the governing equations, which hopefully could be verified in
precision experiments.

Therefore  the expected   impact of the present proposal is twofold.
  As a theoretical result, it goes beyond the original idea of BP  and
constructs
 exact Skyrmions with
intrinsic asymmetry, giving an intriguing generalization of the class of analytic
 functions.  As an application,
our approach constructs  solutions for  Skyrmions, Skyrmion
crystals and Skyrmion lattice with asymmetry, bearing experimental interests in helimagnets.



\begin{thebibliography}{99}

\bibitem{QHexperim}
 M. Lee
 {\it et. al.} , 
Phys. Rev. Lett.    {\bf 102}, 186601 1-4   (2009).
; \ 
 A. Neubauer {\it et. al. }, 
Phys. Rev. Lett.    {\bf 102}, 186602 1-4   (2009).

\bibitem{nutronScattExp}
M\"uhlbauer, S. {\it et. al. }
Science   {\bf 323}, 915   (2009)


 W. M\"unzer {\it et al}, Phys. Rev.  B    {\bf  81}, 041203 (R)  (2010).
\bibitem{LorentzExp} X. Z. Yu et al, Nature       {\bf 465}, 901  (2010).

\bibitem{BP}
 A. A. Belavin  and  A. M.   Polyakov, 
JETP Lett.  {\bf 22},  245    (1975).


 
\bibitem{topological}
 C.  Pfleiderer, S. R.  Julian and 
  G. G. Lonzarich,  
   {
Nature}   {\bf 414}, 427  (2001).

C. Pfleiderer {\it et. al. }
 { 
Nature}   {\bf 427}, 227  (2004)

 
 B. Binz and  A.  Vishwanath,
Physica B   {\bf 403}, 1336   (2008)
\bibitem {top2}

 U. K. R\"o$\beta $ler,  A. N. Bogdanov and  C.
Pfleiderer,
Nature   {\bf 442}, 797   (2006).

U. K. R\"o$\beta$ler, A. A. Leonov, A. N. Bogdanov,  {\it Chiral
Skyrmionic matter in  noncentrosymmetric magnets }, 
arXiv 1009.4849 (2010), U. K. R\"o$\beta$ler, A. A. Leonov, A. N. Bogdanov,
 {\it 
Skyrmion textures  in chiral magnets }, 
arXiv 0907.3651 (2009)

A. B. Butenko,  A. A. Leonov, U. K. R\"o$\beta$ler, A. N. Bogdanov,
Phys. Rev. B {\bf   82}, 052402 (2010)


;


\bibitem {landau}L. D. Landau and E. M. Lifshitz, {\it Statistical Physics} (Pergamon, 
New York), 1977.

\bibitem {dbar} A. S. Fokas, {\it A unified transform method for solving linear and
certain nonlinear PDEs}, Proc. R. Soc. Lond, A   {\bf   453}, 182 (l999)

A.S. Fokas and B. Pelloni, {\it A Transform Method for Evolution PDEs on
the Interval}, IMA J. Appl. Maths {\bf 75} , 564-587 (2005)

B. G. Konopelchenko, {\it On $ bar \partial$-problem and integrable
equations} , ArXiv :nonlin/0002049, (2000)


\bibitem{Moriya} T. Moriya, 
Phys. Rev.  {\bf   120}, 91 (1960)
\bibitem{kundu15b} A. Kundu, {\it Asymmetric Skyrmion 
lattice in helimagnets}, ArXiv :1510.05767  (2015)
\bibitem {ginsperg} P. Ginsparg, {\it
Applied  conformal field theory}, arXiv:hep-th/9108028 (1991)


\end{thebibliography}
\end{document}